\documentclass[12pt]{article}
\usepackage{graphicx}
\usepackage{amssymb}
\usepackage{amsmath}
\usepackage{bm}
\usepackage{cite}
\usepackage{subfigure}

\newcommand{\bear}{\begin{array}}  
\newcommand {\eear}{\end{array}}
\newcommand{\bea}{\begin{eqnarray}}   
\newcommand{\eea}{\end{eqnarray}}
\newcommand{\beq}{\begin{eqnarray}}   
\newcommand{\eeq}{\end{eqnarray}}
\newcommand{\bef}{\begin{figure}}  \newcommand 
{\eef}{\end{figure}}
\newcommand{\bec}{\begin{center}}  \newcommand 
{\eec}{\end{center}}

\newcommand{\Slash}[1]{{\ooalign{\hfil/\hfil\crcr$#1$}}}
\newcommand{\1}{\mbox{1}\hspace{-0.25em}\mbox{l}}

\begin{document}

\begin{titlepage}

\begin{flushright}
IPMU 12-0209
\end{flushright}

\vskip 1.35cm
\begin{center}

{\large 
{\bf 
Neutron Electric Dipole Moment \\ Induced by the Strangeness Revisited
}
}

\vskip 1.2cm

Kaori Fuyuto$^a$,
Junji Hisano$^{a,b}$,
Natsumi Nagata$^{a,c}$

\vskip 0.4cm
{\it $^a$Department of Physics,
Nagoya University, Nagoya 464-8602, Japan}\\
{\it $^b$Kavli IPMU, University of Tokyo, Kashiwa 277-8584, Japan}\\
{\it $^c$Department of Physics, 
University of Tokyo, Tokyo 113-0033, Japan}\\

\date{\today}

\vskip 1.5cm

\begin{abstract} 

 We have revisited the calculation of the neutron electric dipole moment
 in the presence of the CP-violating operators up to dimension five
 based on the chiral perturbation theory. Especially, we focus on the
 contribution of strangeness content. In the calculation, we extract the
 nucleon matrix elements of scalar-type quark operators from the results
 of the lattice QCD simulations, while those of the dipole-type
 quark-gluon operators are evaluated by using the method of the QCD sum
 rules. As a result, it is found that although the strangeness quantity
 in nucleon is small, the contribution of the chromoelectric dipole
 moment of strange quark may be still sizable, and thus may offer
 a sensitive probe for the CP-violating interactions in physics beyond
 the Standard Model. 

\end{abstract}

\end{center}
\end{titlepage}

\section{Introduction}

The neutron electric dipole moment (EDM) is one of the physical
quantities that are extremely sensitive to the CP violation in the
high-energy theories beyond the Standard Model (SM). The
contribution of the SM electroweak interactions to the neutron EDM,
$d_n$, has been evaluated as $d_n\sim 10^{-(31-32)}~e\cdot{\rm cm}$
\cite{Mannel:2012qk, Khriplovich:1981ca}, which is still below the
experimental limit $|d_n|<2.9\times 10^{-26}~e\cdot{\rm cm}$
\cite{Baker:2006ts} by several orders of magnitude. Therefore, the
neutron EDM provides a clean, background-free probe of the CP-violating
interactions in physics beyond the SM, as well as the so-called $\theta$
term in the QCD Lagrangian. 
In order to look for the neutron EDM, various experiments have been
proposed and put into practice. Their sensitivities are expected to be
much improved mainly due to the use of ultra cold neutrons. For
instance, the nEDM collaboration at the Paul-Scherrer
Institute (PSI) \cite{Bodek:2008gr} aims at giving a sensitivity of
$\sim 5\times 10^{-27}~e~{\rm cm}$, and finally getting into the regime
of $10^{-28}~e~{\rm cm}$. 

From a theoretical point of view, on the other
hand, one needs to interpret the value of or the limits on the neutron
EDM provided by experiments in terms of the CP violation at
parton level. To that end, there have been a lot of previous
efforts to derive precise relations between the CP-violating
interactions and the neutron EDM. At the moment, only an approach
exploiting the QCD sum rules \cite{ITEP-73-1978, ITEP-94-1978} provides
a systematic treatment for the derivation, which has been conducted in
various literature \cite{UBCTP-91-21, hep-ph/9705269, hep-ph/9904483,hep-ph/9905317, hep-ph/9908508, hep-ph/0010037, Narison:2008jp,Hisano:2012sc}. In this approach, it is possible to include the contribution of the QCD $\theta$ term, the quark EDMs, and the quark chromoelectric dipole moments (CEDMs) on an equal footing. The QCD lattice simulations also offer a promising way of evaluating the neutron EDM induced by such CP-violating interactions. Although their results
are not precise enough at present \cite{hep-lat/0505022, Berruto:2005hg, lattice_nedm, Shintani:2008nt, Aoki:2008gv}, we expect that the lattice
simulations eventually will be able to determine it with high accuracy.

An alternative, in fact a traditional, approach to the calculation of
the neutron EDM is based on the chiral perturbation theory.  This
method has the advantage of being able to include sea quark effects
indirectly, while in the QCD sum rule method such effects appear only
at the higher orders. The contribution of the QCD $\theta$ term to the
neutron EDM has been evaluated in Refs.~\cite{Crewther:1979pi,
  Pich:1991fq, Borasoy:2000pq, Ottnad:2009jw, Guo:2012vf}. In
Ref.~\cite{Hisano:2004tf, deVries:2010ah, Mereghetti:2010kp}, the quark
CEDM contributions are also 
discussed. While the Peccei-Quinn (PQ)  symmetry \cite{Peccei:1977hh} is
often considered in order to suppress the $\theta$ term and solve the
strong CP problem, it is important to evaluate the neutron EDM induced
by the higher dimensional operators, such as the quark CEDMs, which are
sensitive to TeV-scale physics. In the
calculation, the effects of the CP-violating operators are
incorporated into the CP-odd meson-baryon couplings, which are
obtained by evaluating the nucleon matrix elements of the CP-even
quark scalar operators. The authors in Ref.~\cite{Hisano:2004tf}
extracted the matrix elements from the baryon mass splittings and the
nucleon sigma terms also evaluated in terms of the chiral perturbation
theory, which suggested the sea-quark contribution, especially that of
strange quark, might be considerably large. The recent lattice
simulations, on the other hand, show that the nucleon matrix element
of the strange quark scalar operator is actually quite small contrary
to one's expectation. Since several independent groups have calculated
the values with high accuracy and obtained similar results by using
the different methods, the results have become reliable compared with
the previous estimates. Moreover, a latest calculation based on the
covariant baryon chiral perturbation theory in
Ref.~\cite{Alarcon:2012nr} actually gives a smaller value for the
strangeness content of nucleon than those in the previous studies.
Indeed, it is consistent with the lattice results, while its error is
much larger than those with the lattice simulations.

This situation stimulates the reevaluation of the neutron EDM based on
the chiral perturbation theory with the use of up-to-date results for
the nucleon matrix elements. After the calculation, we will find that
although the strangeness quantity in nucleon is small, the contribution
of the CEDM of strange quark may be sizable, or, even be dominant. 
While the chiral loop computation, especially in the case of the
kaon loop diagrams, might yield large uncertainty, it is found that
similar consequences are achieved, with the cut-off scale of the
loop integral varied within a moderate region. This
result indicates that the strange quark CEDM still may play an important
role in probing the CP-violating interactions in physics beyond the SM.

This paper is organized as follows: we first discuss 
the CP-odd meson-nucleon couplings in the presence of the effective
CP-violating interactions in Sec.~\ref{sec:cpoddpin}. Then, in the
subsequent section, we derive a formulae for the neutron EDM expressed
in terms of the CP-odd couplings. In Sec.~\ref{sec:sqdq}, the nucleon
matrix elements which we use in calculating the CP-odd meson-nucleon
couplings are discussed. Scalar contents of quarks in nucleon are
extracted from the lattice results, while the dipole-type quark-gluon
condensates in nucleon are evaluated by using the method of the QCD sum
rules. The resultant relation between the neutron EDM and the
CP-violating parameters is presented in
Sec.~\ref{sec:results}. Section~\ref{sec:conclusion} is devoted to
conclusion.

\section{CP-odd meson-nucleon couplings}
\label{sec:cpoddpin}

We first write down the effective CP-violating interactions at the scale
of 1 GeV which consist of the flavor-diagonal operators of light quarks
and gluon up to dimension five in QCD and induce the CP-odd
meson-nucleon couplings: 
\begin{eqnarray}
 {\cal L}_{\tiny \Slash{\rm {CP}}}=-\sum_{q=u,d,s}m_q \bar{q}i\theta_q\gamma_5q +
  \theta_G \frac{\alpha_s}{8\pi}G^A_{\mu\nu}\tilde{G}^{A\mu\nu}
-\frac{i}{2}\sum_{q=u,d,s}\tilde{d}_q\bar{q}g_s(G\cdot\sigma)\gamma_5q 
.
\label{Lagrangian}
\end{eqnarray}
Here, $m_q$ are the quark masses, $G^A_{\mu\nu}$ is gluon field strength
tensor, $G\cdot\sigma\equiv
G^A_{\mu\nu}\sigma^{\mu\nu}T^A$, and $\tilde{G}^A_{\mu\nu}\equiv
\frac{1}{2}\epsilon_{\mu\nu\rho\sigma}G^{A\rho\sigma}$ with
$\epsilon^{0123}=+1$. $T^A$ and $g_s$ are the generators and the
coupling constant ($\alpha_s=g_s^2/4\pi)$ of the SU(3)$_{\rm C}$,
respectively. The second term of the above expression is what is called
the effective QCD $\theta$ term, while
the third term represents the chromoelectric dipole moments (CEDMs) for light
quarks. They are dimension-five operators, and thus quite sensitive to
the TeV-scale physics beyond the SM.
The CEDMs of light quarks are not only directly generated by the
CP-violating interactions in the high-energy physics, but also induced
radiatively through the integration of the CP-violating four-quark
operators which include heavy quarks \cite{Hisano:2012cc}.
The coefficients of the
CP-violating operators, $\theta_q$, $\theta_G$, and $\tilde{d}_q$, are
all assumed to be quite small, and we keep only the terms up to the
first order of these parameters.

By using the chiral U(1)$_{\rm A}$ transformation, it is always possible
to rotate out the QCD $\theta$ term into the first term in
Eq.~(\ref{Lagrangian}). In this work, we exploit this basis, where the
effective Lagrangian is given as
\begin{eqnarray}
 {\cal L}_{\tiny \Slash{{\rm CP}}}=-\sum_{q=u,d,s}m_q
  \bar{q}i\theta_q^\prime\gamma_5q  
-\frac{i}{2}\sum_{q=u,d,s}\tilde{d}_q\bar{q}g_s(G\cdot\sigma)\gamma_5q 
.
\label{Lagrangian1}
\end{eqnarray}
with $\sum_{q}\theta_q^\prime =\bar{\theta}\equiv
\theta_G+\sum_{q}\theta_q$. From now on, we only use this basis and omit
the prime, {\it i.e.}, $\sum_{q}\theta_q=\bar{\theta}$. 

We still have some degrees of freedom in the choice of
$\theta_q$ since the SU(3)$_{\rm A}$ chiral rotation transforms a set of
$\theta_q$ into another set. By using the degree of freedom, we take a
basis such that the tad-pole diagrams of the pseudo-scalar mesons should
vanish \cite{Crewther:1979pi}: 
\begin{equation}
 \langle \Omega_{\tiny \Slash{\rm CP}} |{\cal L}_{\tiny \Slash{\rm CP}}
  |\Phi^A\rangle=0~,~~~~(\Phi^A=\pi,~K,~\eta),
  \label{tadpole_conditions}
\end{equation}
where $|\Omega_{\tiny \Slash{\rm CP}}\rangle$ is the vacuum state in the
presence of the CP-violating background sources. By using the partially
conserved axial-vector current (PCAC) relations,
Eq.~(\ref{tadpole_conditions}) leads to the following conditions:
\begin{align}
 \theta_u &=\frac{m_*}{m_u}\biggl[
\bar{\theta}+\frac{m_0^2}{2}\biggl\{
\frac{\tilde{d}_u-\tilde{d}_d}{m_d}+
\frac{\tilde{d}_u-\tilde{d}_s}{m_s}
\biggr\}
\biggr]~, \nonumber \\
 \theta_d &=\frac{m_*}{m_d}\biggl[
\bar{\theta}+\frac{m_0^2}{2}\biggl\{
\frac{\tilde{d}_d-\tilde{d}_u}{m_u}+
\frac{\tilde{d}_d-\tilde{d}_s}{m_s}
\biggr\}
\biggr]~, \nonumber \\
 \theta_s &=\frac{m_*}{m_s}\biggl[
\bar{\theta}+\frac{m_0^2}{2}\biggl\{
\frac{\tilde{d}_s-\tilde{d}_u}{m_u}+
\frac{\tilde{d}_s-\tilde{d}_d}{m_d}
\biggr\}
\biggr]~,
\label{quark_mass_phases}
\end{align}
where
\begin{equation}
 m_*\equiv\frac{m_um_dm_s}{m_um_d+m_dm_s+m_um_s}.
\end{equation}
Also, we parametrize the
condensate $\langle \bar{q}g_s(G\sigma)q\rangle$ as~\cite{Belyaev:1982sa}
\begin{equation}
 \langle \bar{q}g_s(G\sigma)q\rangle =-m_0^2 \langle \bar{q}q\rangle~.
\label{m02def}
\end{equation}

Next, we examine the effects of the CP-violating interactions
on the couplings of baryons with the pseudo-scalar mesons. The couplings
are read off from the pion-baryon scattering amplitude caused by the
CP-violating operators in the low-momentum limit. 
 In Ref.~\cite{Pospelov:2001ys}, the scattering process accompanied by
the creation of an extra pion through the CP-violating interaction is also
considered. The scattering amplitude of the process, however, vanishes
in our calculation thanks to the vacuum alignment condition in
Eq.~(\ref{tadpole_conditions}).
Then, the amplitude is again evaluated by using the PCAC relations as
follows: 
\begin{align}
 \langle \Phi^A B^B|{\cal L}_{\tiny \Slash{\rm CP}}|B^C\rangle
&=\frac{i}{f_\pi}\int d^3x \langle B^B|[J^{A0}_5(0,\bm{x}), {\cal
 L}_{\tiny \Slash{\rm CP}}(0)]|B^C\rangle~,
\label{PCAC}
\end{align}
where $f_\pi\simeq 92.2$~MeV \cite{PDG} is the pion decay constant and
$J^{A\mu}_5(x^0, \bm{x})=\bar{Q}\gamma^\mu \gamma_5 T^AQ$ is the
SU(3)$_{\rm A}$ quark axial vector current with $Q=(u, d, s)^T$ the quark
triplet. The states denoted by $\Phi^A$ and $B^A$
represent the meson and baryon octets, respectively.
We also define the $4\times 4$ matrices ${\cal A}_q$ in the spinor basis
as 
\begin{equation}
 {\cal A}_q\equiv -m_q \theta_q \1 -\frac{1}{2} \tilde{d}_q g_s (G\cdot
\sigma)~,
\end{equation}
and write ${\cal L}_{\tiny \Slash{\rm CP}}$ in the following form:
\begin{equation}
 {\cal L}_{\tiny \Slash{\rm CP}}=\bar{Q}{\cal A}i\gamma_5 Q~,
\end{equation} 
with
\begin{equation}
 {\cal A}={\rm diag}({\cal A}_u, {\cal A}_d, {\cal A}_s),
\end{equation}
being a $3\times 3$ matrix in the flavor basis.
Then, by substituting it into Eq.~(\ref{PCAC}), we obtain the
scattering amplitude as
\begin{equation}
  \langle \Phi^A B^B|{\cal L}_{\tiny \Slash{\rm CP}}|B^C\rangle
=\frac{1}{f_\pi}
 \langle B^B|\bar{Q}\{
T^A,{\cal A}
\}Q|B^C\rangle~.
\label{PCAC1}
\end{equation}
Now all we have to do is to evaluate the baryon matrix elements in the
right-hand side of Eq.~(\ref{PCAC1})~.  
Note that the matrix elements consist of the CP-even operators, though
the interaction we are interested in here is induced by the CP-violating
effects. 

The baryon matrix elements in Eq.~(\ref{PCAC1}) are expressed in terms
of nucleon matrix elements by using the group-theoretical arguments.
A detail discussion is given in Appendix~\ref{baryon_mat}.  For
convenience, we define the following proton matrix elements:
\begin{align}
 S_q\equiv \langle p\vert \overline{q}{q}\vert p\rangle~,
~~~~~~~~
 D_q\equiv \langle p\vert \overline{q}g_s G\cdot \sigma{q}\vert
 p\rangle~.
\label{nucl_mat_def}
\end{align}
Through this paper, the proton state $\vert p(\bm{k})\rangle$ is
normalized as
\begin{equation}
  \langle p(\bm{k}^\prime)\vert p(\bm{k})\rangle
=(2\pi)^3\delta^3(\bm{k}^\prime-\bm{k}).
\end{equation}
Then, using the equations presented in Appendix~\ref{baryon_mat}, we
readily obtain the relations between the CP-odd baryon-meson couplings
and the CP-violating parameters. Among them, we just extract the
couplings which include the neutron field $n$ and charged mesons, since
only such kind of interactions contribute to the neutron EDM. 
They are given as\footnote{
The resultant expressions are consistent with those in
Ref.~\cite{Hisano:2004tf}. }
\begin{equation}
 {\cal L}^{(n)}_{\tiny\Slash{{\rm CP}}}
=\overline{g}_{pn\pi}\overline{p}n\pi^+
+\overline{g}_{\Sigma nK}\overline{\Sigma^-}nK^-+{\rm h.c.}~,
\label{CPodd1}
\end{equation}
with
\begin{align}
 \overline{g}_{pn\pi}&=\frac{1}{3\sqrt{2}f_\pi}\bigl(2X_1+
X_8\bigr)\nonumber \\
&=-\frac{1}{ \sqrt{2}f_\pi}\bigl[
(m_u\theta_u+m_d\theta_d)(S_u-S_d)+\frac{1}{2}
(\tilde{d}_u+\tilde{d}_d)(D_u-D_d)
\bigr],\nonumber \\
\overline{g}_{\Sigma nK}&=\frac{1}{6\sqrt{2}f_\pi}\bigl(4Y_1
+3Y_3-Y_8\bigr)\nonumber \\
&=-\frac{1}{ \sqrt{2}f_\pi}\bigl[
(m_u\theta_u+m_s\theta_s)(S_s-S_d)+\frac{1}{2}
(\tilde{d}_u+\tilde{d}_s)(D_s-D_d)
\bigr],
\label{CPodd2}
\end{align}
where $X_i$ and $Y_i$ are defined in Appendix.~\ref{baryon_mat}.
Thus, evaluation of the CP-odd baryon-meson couplings reduces to that of
the proton matrix elements, {\it i.e.}, $S_q$ and $D_q$. 

Of particular interest is the case where the PQ symmetry
\cite{Peccei:1977hh} is imposed. In such a case, the expressions presented
above are modified. Since there exist other CP-violating sources than the QCD
$\theta$ term, $\bar{\theta}$ is not completely erased by the PQ mechanism but
effectively induced as \cite{Bigi:1991rh, Pospelov:1997uv}
\begin{equation}
 \bar{\theta}_{\rm
  PQ}=\frac{m_0^2}{2}\sum_{q=u,d,s}\frac{\tilde{d}_q}{m_q} ~.
\end{equation}
Then, the CP-odd baryon-meson couplings lead to
\begin{align}
 \overline{g}_{pn\pi}\vert_{\rm PQ}&=-\frac{1}{\sqrt{2}f_\pi}[
\frac{m_0^2}{2}(S_u-S_d)+\frac{1}{2}(D_u-D_d)](\tilde{d}_u+\tilde{d}_d),
\nonumber \\
 \overline{g}_{\Sigma nK}\vert_{\rm PQ}&=-\frac{1}{\sqrt{2}f_\pi}[
\frac{m_0^2}{2}(S_s-S_d)+\frac{1}{2}(D_s-D_d)](\tilde{d}_u+\tilde{d}_s).
\label{CPoddPQ}
\end{align}
Note that in both Eqs.~(\ref{CPodd2}) and (\ref{CPoddPQ}), the
contribution of the strange quark CEDM, $\tilde{d}_s$, does not
necessarily vanish\footnote{
On the contrary, the strange CEDM contribution to the isospin-conserving
nucleon-$\eta^0$ coupling is suppressed in the absence of the strangeness
content in nucleon. 
} even if the strange quark content in nucleon is quite
small, {\it i.e.}, $S_s,~D_s\to 0$. This observation allows us to expect
that the contribution of strangeness to the neutron EDM is sizable even
in such a case, and it will be actually shown in the following
discussion.

\section{Neutron Electric Dipole Moment}
\label{sec:nEDM}

Before evaluating the proton matrix elements, we deduce a formula for
the neutron EDM induced by the CP-odd baryon-meson couplings given in
the previous section. To that end, we first obtain the CP-even vertices
which we will exploit to calculate the neutron EDM. Such interactions
are included in the following effective Lagrangian:
\begin{align}
 {\cal L}^0&=
\frac{f_\pi^2}{4}{\rm Tr}[D_\mu U (D^\mu U)^\dagger]+
{\rm Tr}[\overline{B}(i\Slash{D}-M_{\rm B})B]
\nonumber \\
&-\frac{D}{2}{\rm Tr}\bigl(\overline{B}\gamma^\mu\gamma_5\{
\xi_\mu, B\}\bigr)
-\frac{F}{2}{\rm Tr}\bigl(\overline{B}\gamma^\mu \gamma_5
[\xi_\mu , B]\bigr),
\label{L0}
\end{align}
with appropriate mass terms for the meson/baryon mass spectrum.
Here, $U$ is defined as $U\equiv \xi^2$
with
\begin{equation}
 \xi=\exp\biggl(
\frac{i\Phi}{\sqrt{2}f_\pi}
\biggr),
\end{equation}
and
\begin{equation}
 \Phi=\sqrt{2}\Phi^AT^A=
\begin{pmatrix}
  \frac{\pi^0}{\sqrt{2}}+\frac{\eta^0}{\sqrt{6}}& \pi^+& K^+\\
  \pi^- &  -\frac{\pi^0}{\sqrt{2}}+\frac{\eta^0}{\sqrt{6}}& K^0\\
  K^- & \overline{K^0}& -2\frac{\eta^0}{\sqrt{6}}
\end{pmatrix}
.
\end{equation}
The baryon matrix field $B$ is defined by
\begin{eqnarray}
 B=\frac{1}{\sqrt{2}}B_a\lambda_a
=
\begin{pmatrix}
 \frac{\Sigma^0}{\sqrt{2}}+\frac{\Lambda^0}{\sqrt{6}}& \Sigma^+& p\\
  \Sigma^- &  -\frac{\Sigma^0}{\sqrt{2}}+\frac{\Lambda^0}{\sqrt{6}}& n\\
  \Xi^- & {\Xi^0}& -2\frac{\Lambda^0}{\sqrt{6}}
\end{pmatrix}
~,
\label{baryon_mat_field}
\end{eqnarray}
with $\lambda_a$ the Gell-Mann matrices. The covariant derivatives in
Eq.(\ref{L0}) are given as 
\begin{align}
 D_\mu U&\equiv \partial_\mu U -ir_\mu U+iUl_\mu~,
\nonumber \\
 D_\mu B&\equiv \partial_\mu B+[\Gamma_\mu, B],
\end{align}
where
\begin{equation}
 \Gamma_\mu\equiv\frac{1}{2}[\xi^\dagger (\partial_\mu-i r_\mu)\xi
+\xi(\partial_\mu -il_\mu)\xi^\dagger],
\label{chiral_connection}
\end{equation}
is called the chiral connection,
and $r_\mu$ and $l_\mu$ denote the external sources for the right- and
left-handed currents, respectively. Further, the definition of $\xi_\mu$
is 
\begin{equation}
 \xi_\mu \equiv i[\xi^\dagger (\partial_\mu -ir_\mu )\xi
-\xi(\partial_\mu -il_\mu )\xi^\dagger],
\end{equation}
which is referred to as the chiral vielbein.
The low-energy constants $D$ and $F$ are determined by fitting the
semi-leptonic decays $B\to B^\prime +e^-+\bar{\nu}_e$ at tree level
\cite{Hsueh:1988ar}: $D=0.80(1), ~F=0.47(1)$.
From Eq.~(\ref{L0}), we extract the CP-even meson-baryon
interactions. They are expressed as
 \begin{equation}
 {\cal L}_{\pi BB}=\frac{1}{f_\pi}(d_{ABC}D-i f_{ABC}F) \partial_\mu
\pi^A \overline{B}^B\gamma^\mu \gamma_5 B^C,
\end{equation}
with $f_{ABC}$ the structure constant of $SU(3)$ and $d_{ABC}$ defined
by $d_{ABC}\equiv 2{\rm Tr}[\{T^A, T^B\}T^C]$. Among the terms, 
\begin{equation}
 {\cal L}_{\rm CP~ even}^{(n)}
=\frac{1}{\sqrt{2} f_\pi}(D+F)\overline{n}\gamma^\mu \gamma_5
p\partial_\mu\pi^- 
+\frac{1}{\sqrt{2} f_\pi}(D-F)\overline{\Sigma^-}\gamma^\mu \gamma_5 n
\partial_\mu K^-
+{\rm h.c.}
\end{equation}
contributes to the neutron EDM. Further, by setting $r_\mu = l_\mu
=-eQA_\mu$ with $e$ the electric charge of positron, {\it i.e.}, $e>0$,
and $Q={\rm diag}(2/3,-1/3,-1/3)$, 
we obtain the interactions of photon with mesons and baryons. The
relevant terms are as follows:
\begin{align}
 {\cal L}_\gamma=&
-ie A_\mu (\partial^\mu \pi^+ \pi^- -\pi^+ \partial^\mu \pi^-
+\partial^\mu K^+K^- -K^+\partial^\mu K^-)\nonumber \\
&-e(\overline{p}\Slash{A}p+\overline{\Sigma^+}\Slash{A}\Sigma^+
 -\overline{\Sigma^-} \Slash{A}\Sigma^-)
\nonumber \\
&+\frac{ie}{\sqrt{2} f_\pi}(D+F)\overline{p}\Slash{A}\gamma_5 n\pi^+
-\frac{ie}{\sqrt{2} f_\pi}(D-F)\overline{\Sigma^-}\Slash{A}\gamma_5
 nK^-
+{\rm h.c.}.
\end{align}
Here, $A_\mu$ denotes the electromagnetic field.

\begin{figure}[t]
\begin{center}
\includegraphics[height=3cm]{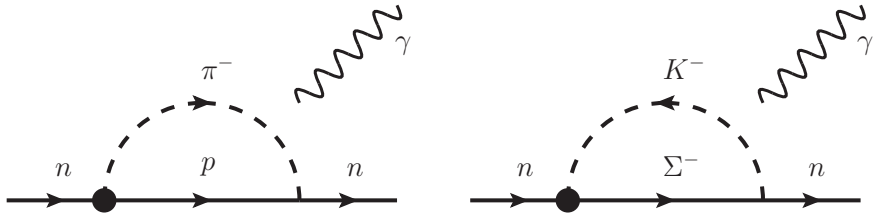}
\caption{Diagrams which give rise to the neutron EDM. Solid, dashed, and
 wavy lines represent baryons, mesons, and photon, respectively. 
 Further, dots indicate the CP-odd
 interactions given in Eqs.~(\ref{CPodd1}) and (\ref{CPodd2}).} 
\label{fig:loop}
\end{center}
\end{figure}

By using the interactions obtained above, we now evaluate the neutron
EDM. It is induced by the diagrams displayed in
Fig.~\ref{fig:loop}. Here, solid, dashed, and wavy lines represent baryons,
mesons, and photon, respectively.
 Further, dots indicate the CP-odd interactions given in
Eqs.~(\ref{CPodd1}) and (\ref{CPodd2}). The resultant formula\footnote{
Note that the formula presented here is different from that in
Ref.~\cite{Hisano:2004tf} by a factor of two.
} for the
neutron EDM is as follows:
\begin{equation}
 d_n=-\frac{e}{4\sqrt{2}\pi^2f_\pi}\biggl[
(D+F)\overline{g}_{pn\pi}\log\biggl(\frac{\Lambda}{m_\pi}\biggr)
-(D-F)\overline{g}_{\Sigma nK}\log\biggl(\frac{\Lambda}{m_K}\biggr)
\biggr]~,
\label{dn_result}
\end{equation}
with $m_\pi$ and $m_K$ the masses of pion and kaon,
respectively. Here, we extract the terms which contain logarithmic
singularities in the chiral limit. $\Lambda\simeq M_{\rm B}$ is a UV
cutoff.  Furthermore, there exists tree-level, higher order
contribution which acts as counterterms \cite{Pich:1991fq}. With the
contribution added to the above equation, we obtain scale-independent
formula for the neutron EDM. However, the low-energy constants for the
operators which give rise to the tree-level contribution are not
determined through the symmetry arguments. Thus, these unknown factors
as well as the lack of other non-singular terms in Eq.~(\ref{dn_result})
result in theoretical uncertainty. In the following calculation, we
just directly use the expression (\ref{dn_result}) with $\Lambda$
taken to be the proton mass as an approximation. 
Note that the size of $\log(\Lambda/m_K)$ is not large enough and
thus might be considerably suffered from the uncertainty coming from the
counterterms.
The error originating from the approximation is to be estimated by
varying the cutoff scale, similar to Ref.~\cite{Pich:1991fq}.

\section{Evaluation of $S_q$ and $D_q$}
\label{sec:sqdq}

Now all we have to do is to evaluate the proton matrix elements, $S_q$
and $D_q$. 

\subsection{$S_q$}

As we mentioned in the Introduction, $S_q$ is directly extracted
from the lattice QCD simulations. 
To evaluate them, we first introduce the following parameters:
\begin{eqnarray}
\sigma_{\pi N}&=&\frac{m_u+m_d}{2} \langle p|\bar{u}u+\bar{d}d|p\rangle  \ 
,\nonumber\\
\sigma_s&=&{m_s} \langle p|\bar{s}s|p\rangle  \ 
,\nonumber\\
\xi&=&\frac{\langle p|\bar{u}u-\bar{d}d|p\rangle}
{\langle p|\bar{u}u+\bar{d}d|p\rangle}\ .
\end{eqnarray}
Exploiting these parameters, we express $S_q$ as
\begin{align}
 S_u&=\frac{1}{2}\frac{2}{(m_u+m_d)}(1+\xi)\sigma_{\pi
 N},~\nonumber \\
 S_d&=\frac{1}{2}\frac{2}{(m_u+m_d)}(1-\xi)\sigma_{\pi
 N},~\nonumber \\
 S_s&=\frac{\sigma_s}{m_s}.
\end{align}
Recent lattice simulations predict \cite{Aoki:2008sm, Shanahan:2012wh}
\begin{align}
 \sigma_{\pi N}&=45\pm 6~{\rm MeV},\nonumber \\
 \sigma_s&=21\pm 6~{\rm MeV}.
\end{align}
The symmetry breaking parameter $\xi$ is obtained from the baryon mass
splittings as \cite{Cheng:1988im}
\begin{align}
 \xi&=\frac{m_{\Xi^0}+m_{\Xi^-} -m_{\Sigma^+}-m_{\Sigma^-}}
{m_{\Xi^0}+m_{\Xi^-} +m_{\Sigma^+}+m_{\Sigma^-}-2m_p-2m_n}
\biggl(
1-\frac{m_u+m_d}{2m_s}\frac{2\sigma_s}{\sigma_{\pi N}}
\biggr) \nonumber \\
&=0.197 \biggl(
1-\frac{m_u+m_d}{2m_s}\frac{2\sigma_s}{\sigma_{\pi N}}
\biggr)~.
\end{align}
Then, with the quark-mass parameters \cite{PDG}
\begin{equation}
 \frac{m_u+m_d}{2}=4.6\pm 0.7~{\rm MeV},~~~~~~
  m_s =115\pm 6~{\rm MeV},
\end{equation}
we obtain the values of $S_q$ as
\begin{align}
 S_u&=5.8,~~~~~~
 S_d=4.0,~~~~~~
 S_s=0.18.
\label{Sq_result}
\end{align}
Here, we evolve the masses of light quarks from the scale
$\mu=2~{\rm GeV}$ to $\mu=1~{\rm GeV}$ by using the renormalization
group equations.
Note that the values of $S_u$ and $S_d$ are almost as twice as those in
Ref.~\cite{Hisano:2004tf}. This is mainly due to the quark-mass
parameters. Moreover, $S_s$ is much smaller than that in
Ref.~\cite{Hisano:2004tf}, since recent lattice simulations predict
relatively small values for $\sigma_s$. This fact drives us to
reevaluate the strange quark contribution to the neutron EDM, as we
mentioned to in the Introduction.

\subsection{$D_q$}
\label{Dq}

For the values of $D_q$, on the other hand, we are not able to use the
lattice simulations at the present time. Although it is desirable that
the lattice simulations eventually determine $D_q$, for the time being,
we evaluate them by using the method of QCD sum rules
\cite{ITEP-73-1978, ITEP-94-1978}. A similar approach is studied in
Ref.~\cite{Pospelov:2001ys}. 

For this purpose, we first add to the ordinary
QCD Lagrangian the following terms which couple to the external
$c$-number fields $J_q$:
\begin{equation}
 {\cal L}_{\rm ext}=J_q\bar{q}g_sG\cdot \sigma q ~.
\label{external}
\end{equation}
The method of QCD sum rules requires that we evaluate the
correlation function of the proton interpolating fields in terms of two
different ways; one is to describe it from the phenomenological point
of view, and the other is to calculate it in terms of the operator
product expansion (OPE). The correlation function we are to evaluate is
defined as follows:
\begin{equation}
 \Pi(k)\vert_J\equiv i\int d^4 xe^{ik\cdot x}\langle \Omega |
T\{\eta_p(x), \overline{\eta}_p(0)\}|\Omega\rangle_{J}~,
\label{correlator}
\end{equation}
with $\eta_p(x)$ the interpolating field for proton. The subscript of
the correlator, $J$, indicates that the function in
Eq.~(\ref{correlator}) is evaluated in the presence of external fields.
We expand the correlation function in terms of $J_q$ as
\begin{equation}
 \Pi(k)\vert_J=\Pi^0(k)+J_q\Pi_q(k)+{\cal O}(J^2),
\end{equation}
and focus on the correlator $\Pi_q$ since it includes information on
$D_q$.

The matrix element of the proton interpolating field between the vacuum
and the one-particle proton state is given as
\begin{equation}
 \langle \Omega |\eta_p(x)|p(\bm{k}, s)\rangle
=\frac{\lambda_p}{\sqrt{2m_p}} u_p(\bm{k},s)e^{-ik\cdot x}~.
\end{equation}
Here, $u_p(\bm{k},s)$ is a spinor wave function, which is normalized as
usual, {\it i.e.}, $\bar{u}(\bm{k},s^\prime)u(\bm{k},s)=2m_p
\delta_{ss^\prime}$. We parametrize the field 
renormalization constant as $\lambda_p$, whose value is to be determined
later.

A phenomenological description of the correlator is readily given as 
\begin{align}
 \Pi_q(k)=\lambda_p^2 D_q 
  \frac{1}{\Slash{k}-m_p}  \frac{1}{\Slash{k}-m_p}
 +\dots~,
\end{align}
where the dots indicate the contribution of the excited states. Among
the terms in the above expression, we hereafter focus on the terms that
contain only one gamma matrix, that is, the terms proportional to
$\Slash{k}$. Then, it follows that
\begin{align}
 \Pi_q(k)=\Slash{k} f_q(k^2)+\dots~,
\label{phen_corr}
\end{align}
with
\begin{align}
 f_q(k^2)\equiv
  \frac{2m_p\lambda_p^2D_q}{(k^2-m_p^2)^2}
+\frac{A_q(k^2)}{k^2-m_p^2} +B_q(k^2)~.
\end{align}
Here, $A_q(k^2)$ and $B_q(k^2)$ are the functions that have no pole at
$k^2=m_p^2$. Dots in Eq.~(\ref{phen_corr}) correspond to terms with no
gamma matrices. 

In order to evaluate the correlation function by using the OPE, on
the other hand, one needs to express the proton interpolating field in
terms of a composite operator of quarks which has the same quantum
numbers as those of a proton. The general form of the proton
interpolating field is given as
\begin{equation}
 \eta_p(x)=j_{p1}(x)+\beta j_{p2}(x)~,
\end{equation}
where
\begin{align}
 j_{p1}(x)&\equiv 2\epsilon^{abc}[u^T_a(x)C\gamma_5 d_b(x)]u_c(x),
\nonumber  \\
 j_{p2}(x)&\equiv 2\epsilon^{abc}[u^T_a(x)C d_b(x)]\gamma_5u_c(x).
\end{align}
Here the subscripts, $a,b,c$, are the color indices and $C$ denotes the
charge conjugation matrix. Although the interpolator $j_{p2}(x)$ vanishes
in the non-relativistic limit, there is no reason for the interpolator
to be excluded since light quarks in a proton are actually
relativistic. The unphysical parameter $\beta$ is to be fixed later.
The OPE calculation is conducted with the quark propagators and
condensates in the presence of the external fields. They are derived in
Appendix~\ref{propagator_correlator}. By using them, we evaluate the
correlation function and extract the terms proportional to the external
fields. First, we decompose the correlation function as
\begin{align}
 \Pi(x)|_J&=  \langle \Omega|T\{\eta_p(x)\bar{\eta}_p(0)\}|
  \Omega\rangle_{J} \nonumber\\
&= \langle \Omega|T\{j_{p1}(x)\bar{j}_{p1}(0)\}|
  \Omega\rangle_{J}
+\beta\{\langle \Omega|T\{j_{p2}(x)\bar{j}_{p1}(0)\}|
  \Omega\rangle_{J}
\nonumber \\
&~~~~~~~~~~~~~
+\langle \Omega|T\{j_{p1}(x)\bar{j}_{p2}(0)\}|
  \Omega\rangle_{J}\}
+\beta^2\langle \Omega|T\{j_{p2}(x)\bar{j}_{p2}(0)\}|
  \Omega\rangle_{J}~.
\end{align}
For convenience, we use the following abbreviation:
\begin{equation}
  \Pi_{kl}(x)=  
\langle \Omega|T\{j_{pk}(x)\bar{j}_{pl}(0)\}|
  \Omega\rangle~,~~~~~~(k,l=1,2) ~.
\end{equation}
They are expressed in terms of the propagators $S^q_{ab}(x)$ given in
Eq.~(\ref{propagator}) in Appendix.~\ref{propagator_correlator} as
follows: 
\begin{align}
 \Pi_{11}(x)&=4\epsilon_{abc}\epsilon_{a^\prime b^\prime c^\prime}
\bigl[
{\rm Tr}\{
\overline{S}^u_{aa^\prime}(x)\gamma_5 S^d_{bb^\prime}(x)\gamma_5
\}
S^u_{cc^\prime}(x)
+S^u_{aa^\prime}(x)\gamma_5\overline{S}^d_{bb^\prime}(x)
\gamma_5S^u_{cc^\prime}(x)
\bigr],\nonumber \\
 \Pi_{12}(x)&=4\epsilon_{abc}\epsilon_{a^\prime b^\prime c^\prime}
\bigl[
{\rm Tr}\{
\overline{S}^u_{aa^\prime}(x)\gamma_5 S^d_{bb^\prime}(x)
\}
S^u_{cc^\prime}(x)\gamma_5
+S^u_{aa^\prime}(x)\overline{S}^d_{bb^\prime}(x)
\gamma_5S^u_{cc^\prime}(x)\gamma_5
\bigr],\nonumber \\
 \Pi_{21}(x)&=4\epsilon_{abc}\epsilon_{a^\prime b^\prime c^\prime}
\bigl[
{\rm Tr}\{\gamma_5
\overline{S}^u_{aa^\prime}(x) S^d_{bb^\prime}(x)
\}
\gamma_5S^u_{cc^\prime}(x)
+\gamma_5S^u_{aa^\prime}(x)\gamma_5\overline{S}^d_{bb^\prime}(x)
S^u_{cc^\prime}(x)
\bigr],\nonumber \\
 \Pi_{22}(x)&=4\epsilon_{abc}\epsilon_{a^\prime b^\prime c^\prime}
\bigl[
{\rm Tr}\{
\overline{S}^u_{aa^\prime}(x) S^d_{bb^\prime}(x)
\}\gamma_5
S^u_{cc^\prime}(x)\gamma_5
+\gamma_5S^u_{aa^\prime}(x)\overline{S}^d_{bb^\prime}(x)
S^u_{cc^\prime}(x)\gamma_5
\bigr],
\end{align}
where
\begin{equation}
 \overline{S}^q_{ab}(x)\equiv
CS^{qT}_{ab}(x)C^\dagger.
\end{equation}
Then, a series of Wick contraction leads to 
\begin{align}
 \Pi_{11}(x)\vert_J&=-\frac{i}{8\pi^4}\langle\bar{q}q\rangle
\frac{\Slash{x}}{(x^2)^3}(9J_u+5J_d)m_0^2, \nonumber \\
 \Pi_{12}(x)\vert_J&=-\frac{i}{8\pi^4}\langle\bar{q}q\rangle
\frac{\Slash{x}}{(x^2)^3}(3J_u+J_d)m_0^2, \nonumber \\
 \Pi_{21}(x)\vert_J&=-\frac{i}{8\pi^4}\langle\bar{q}q\rangle
\frac{\Slash{x}}{(x^2)^3}
(3J_u+J_d)m_0^2, \nonumber \\
 \Pi_{22}(x)\vert_J&=-\frac{i}{8\pi^4}\langle\bar{q}q\rangle
\frac{\Slash{x}}{(x^2)^3}(9J_u+5J_d)m_0^2,
\end{align}
up to the leading order. Here, we keep only the terms including an
external field. It is followed that
\begin{align}
 \Pi(x)\vert_J=-\frac{i}{8\pi^4}\langle\bar{q}q\rangle
\frac{\Slash{x}}{(x^2)^3}m_0^2\bigl[(9\beta^2+6\beta+9)J_u+
(5\beta^2+2\beta+5)J_d
\bigr],
\end{align}
and its Fourier transformation results in
\begin{align}
 \Pi(k)\vert_J=-\frac{1}{32\pi^2}\langle\bar{q}q\rangle
\Slash{k}\log\biggl(
\frac{-k^2}{\Lambda^2}
\biggr)m^2_0\bigl[(9\beta^2+6\beta+9)J_u+(5\beta^2+2\beta+5)J_d
\bigr],
\label{result_corr}
\end{align}
with $\Lambda$ a certain ultraviolet mass scale.

Then, by applying the Borel transformation to Eqs.~(\ref{phen_corr}) and
(\ref{result_corr}), we now obtain the sum rules for $D_q$ as
\begin{align}
 D_u-\frac{A_u}{2m_p\lambda_p^2}M^2&
=+\frac{M^4}{32\pi^2}\frac{\langle \bar{q}q\rangle}{2m_p\lambda_p^2}
\exp\biggl\{\frac{m_p^2}{M^2}\biggr\}(9\beta^2+6\beta+9)m^2_0,
 \nonumber \\
 D_d-\frac{A_d}{2m_p\lambda_p^2}M^2&
=+\frac{M^4}{32\pi^2}\frac{\langle \bar{q}q\rangle}{2m_p\lambda_p^2}
\exp\biggl\{\frac{m_p^2}{M^2}\biggr\}(5\beta^2+2\beta+5)m_0^2,
\label{sum_rules}
\end{align}
and $D_s=0$ up to the leading order calculation. When we drive the
sum rules, we assume that the single-pole contributions
$A_q$ scarcely depend on $k^2$ around $k^2=m_p^2$, and regard them as
constants. Further, we just neglect $B_q$ with expecting that their
contributions are sufficiently reduced by the Borel transformation. Note
that the sum rules in Eq.~(\ref{sum_rules}) contain the identical
function of $M^2$, $M^4\exp(m_p^2/M^2)$, in their right-hand
side. Moreover, the tangent line to the function at a given Borel mass
squared $M^2$ gives us both the first and second terms in the left-hand
side of the sum rules~\cite{Hisano:2012sc}. Especially, if one sets the
Borel mass at the minimal of the function, {\it i.e.}, $M^2=m_p^2/2$,
the single-pole contributions vanish. Since this choice also makes the
sum rules stable under the slight variation of the Borel mass, we adopt
it in the following discussion.

Now all we have to do reduces to the determination of the constant
$\lambda_p$. One way to evaluate it is again using the method of QCD sum
rules. In Ref.~\cite{Leinweber:1995fn}, two sum rules which allow us to
extract $\lambda_p$ are presented; one contains the Lorentz
structure $\Slash{p}$ and the other is proportional to unity. On the
other hand, it is possible to extract the value from the lattice
result. By using the parameters obtained in Ref.~\cite{Aoki:2008ku}, we
obtain
\begin{equation}
 \lambda_p=2\times 0.91\cdot\bigl[(\alpha_1-\alpha_2)- \beta
  (\alpha_1+ \alpha_2)\bigr]~,
\label{lambda_gen_lat}
\end{equation}
with the parameters $\alpha_1$ and $\alpha_2$ given as
\begin{equation}
 \alpha_1=-0.0112\pm 0.0012_{(\rm stat)} \pm0.0022_{(\rm syst)}~{\rm
  GeV}^3~,
\nonumber
\end{equation}
\begin{equation}
 \alpha_2=0.0120\pm 0.0013_{(\rm stat)} \pm0.0023_{(\rm syst)}~{\rm GeV}^3~.
\end{equation}
Here we take the renormalization effect into account
\cite{Hisano:2012sc}. These values are consistent with those
obtained by another group \cite{Braun:2008ur} within the errors of their
simulations.
In the following calculation, we use the values
for $\lambda_p$ evaluated in both approaches, and present two results
corresponding to these two values.

The estimation of $D_q$ presented here is, however,
to be considered as just for reference. At the moment we only
present center values for those quantity. In order to estimate the
theoretical error for the calculation and to improve the accuracy of the
result, we need the investigation of the excited/continuum state
contribution as well as the execution of the higher order calculation
for the OPE. In the higher order calculation, there exist several unknown
condensates such as the susceptibilities of $\langle \bar{q}q \rangle$
with respect to the external sources, and thus they make it difficult
to give a precise prediction. Therefore, much detail analysis with the
determination of these unknown condensates is required for a robust
calculation of $D_q$. We hope our study presented in this paper
stimulates a lot of efforts to evaluate the quantity with various
methods, especially with the lattice QCD simulations.

\section{Results}
\label{sec:results}

Now we derive a relation between the neutron EDM with the CP-violating
parameters by using the nucleon matrix elements obtained above. We use
the values in Eq.~(\ref{Sq_result}) for $S_q$, while $D_q$ ($q=u,d$) are
evaluated from the sum rules in Eq.~(\ref{sum_rules}) with $D_s$ presumed
to be zero. 

We are interested in the significance of the strange CEDM contribution
to the neutron EDM. In order to examine it, we consider each
contribution of CEDM in the presence of PQ symmetry. In this case, the
resultant relation between the CEDMs and the neutron EDM is expressed as
\begin{equation}
 d_n\vert_{\rm PQ}=\sum_{q}eC_q\frac{\tilde{d}_q}{m_q}.
\end{equation}
Here, we normalize the CEDMs by the quark masses, since in most cases the
quark CEDMs are proportional to the quark masses, $\tilde{d}_q\propto
m_q$. Then, we look into the ratio of the coefficient $C_s$ 
against those of up and down quarks. It allows us to evade an
overall factor of uncertainty coming from the logarithmic factor in
Eq.~(\ref{dn_result}), though the relative values of $C_q$ still might
be affected.
\begin{figure}[t]
\begin{center}
\includegraphics[height=7cm]{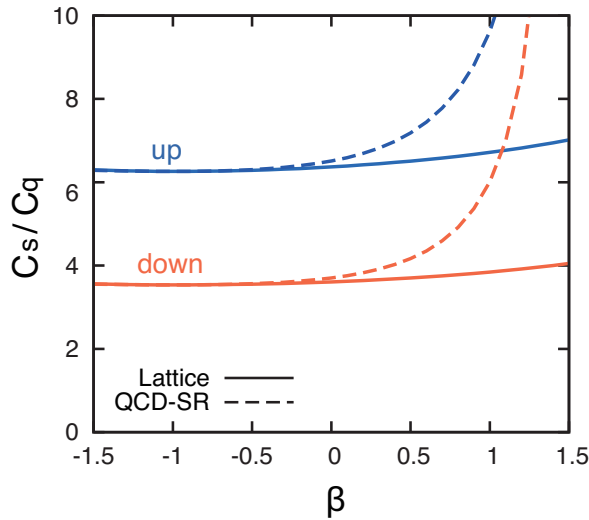}
\caption{Ratios $C_s/C_q$ ($q=u,d$) as functions of the
parameter $\beta$ in the sum rules. Upper blue (lower orange) lines show
 $C_s/C_u$ ($C_s/C_d$). Solid lines
correspond to the results where we use $\lambda_p$ in
Eq.~(\ref{lambda_gen_lat}), while dashed lines represent those
obtained with $\lambda_p$ evaluated in Ref.~\cite{Leinweber:1995fn}.} 
\label{fig:beta}
\end{center}
\end{figure}

In Fig.~\ref{fig:beta}, we plot the ratios $C_s/C_q$ as functions of the
parameter $\beta$ in the sum rules. The upper blue and the lower orange
lines show $C_s/C_u$ and $C_s/C_d$, respectively. The solid lines
correspond to the results in which we use $\lambda_p$ in
Eq.~(\ref{lambda_gen_lat}), while the dashed lines represent those
obtained with $\lambda_p$ evaluated in Ref.~\cite{Leinweber:1995fn}.
In the calculation, we take $m_0^2=0.8~{\rm GeV}^2$
\cite{Belyaev:1982sa} and $\langle \bar{q}q\rangle$ is evaluated using
the relation $\langle \bar{q}q\rangle=-m_{\pi}^2f_{\pi}^2/(m_u+m_d)$ as
$\langle \bar{q}q\rangle\simeq-(262~{\rm MeV})^3$ at the scale of
$\mu= 1~{\rm GeV}$. In addition, when we
compute $\lambda_p$ from the results in Ref.~\cite{Leinweber:1995fn}, we
use the sum rule for $\Slash{p}$ and set the threshold $w$ and the gluon
condensate $b\equiv(2\pi)^2(\frac{\alpha_s}{\pi}GG)$ as $w=2~{\rm GeV}$
and $b=470~{\rm MeV}^4$, respectively. With the parameters we
obtain $\vert \lambda_p\vert\simeq 0.04~{\rm GeV}^3$ for $\beta=-1$ and
$\vert \lambda_p\vert\simeq 0.02~{\rm GeV}^3$ for $\beta=+1$.
It is found that both
calculations for $\lambda_p$ give rise to similar results in the case of
$\beta\lesssim0$, while one deviates form the other for positive
$\beta$. In any case, the strange CEDM contribution is sizable, in fact
dominates the other contributions, though the strangeness content
of nucleon is quite small. The contribution comes from the $K$-meson
loop and does not vanish in the limit of small $S_s$ and $D_s$, as it is
understood from Eq.~(\ref{CPoddPQ}).\footnote{The strange CEDM
contribution to the neutron EDM vanishes if $m_0^2 S_d+D_d=0$, though it
looks accidental.}

\begin{figure}[t]
\begin{center}
\includegraphics[height=7cm]{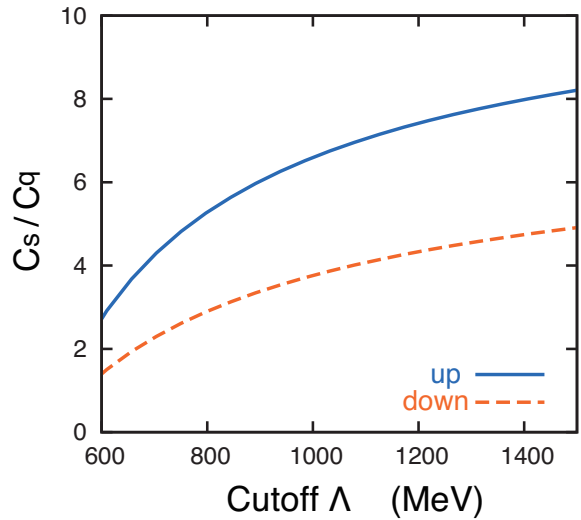}
\caption{Dependence of $C_s/C_q$ on cutoff scale $\Lambda$. Blue solid
 and orange dashed lines show the results for up and down quarks,
 respectively. We use $\lambda_p$ with $\beta=-1$ in
 Eq.~(\ref{lambda_gen_lat}) in the calculation. } 
\label{fig:cutoff}
\end{center}
\end{figure}

As we have mentioned to above, the computation suffers from the
uncertainty due to the tree-level contribution of the unknown
counterterms as well as the lack of other non-singular terms in
Eq.~(\ref{dn_result}). To estimate its significance, we vary the cutoff
scale 
$\Lambda$ in Eq.~(\ref{dn_result}), following the analysis in
Ref.~\cite{hep-ph/0010037}. In Fig.~\ref{fig:cutoff}, we plot
the dependence of the ratios $C_s/C_q$ on $\Lambda$. 
The blue solid (orange dashed)
line shows the result for $C_s/C_u$ ($C_s/C_d$). We use the lattice
result in Eq.~(\ref{lambda_gen_lat}) for the value of $\lambda_p$ with
$\beta$ set to be $-1$. This
figure tells us that although the variation of the cutoff scale may
change the ratio by an ${\cal O}(1)$ factor, the contribution of strange
CEDM tends to remain important.
Taking the results obtained above into consideration, we conclude that
although the recent lattice results indicate that the 
amount of strange quark in nucleon is small, the CEDM of strange quark
may play an important role in probing the CP-violating interactions in
physics beyond the SM with the neutron EDM. 

Finally, as a reference value, we present the numerical results for the
relation between the CEDMs and the neutron EDM:
\begin{equation}
 d_n=e(5.8\times 10^{-16}\bar{\theta}~[{\rm
  cm}]-1.4\tilde{d}_u+0.7\tilde{d}_d+0.4\tilde{d}_s),
\label{dn}
\end{equation}
where we take $\beta=-1$ and use $\lambda_p$ in
Eq.~(\ref{lambda_gen_lat}). Note that unlike the CEDM contribution, the
calculation of the $\bar{\theta}$ contribution is not affected by the
uncertainty in the values of $D_q$, since the contribution only depends
on $S_q$. In this sense, the result for the $\bar{\theta}$ contribution
to the neutron EDM is robust up to the uncertainty coming from the
chiral loop calculation.
In the presence of the PQ
symmetry, on the other hand, the relation is modified to 
\begin{align}
 d_n\vert_{\rm PQ}=e(3.0\tilde{d}_u+2.5\tilde{d}_d+0.5\tilde{d}_s).
\label{dnPQ}
\end{align}
Here, we would like to remark that there may be large error
in estimating the size of the strange CEDM contribution, which results
from the uncertainty of the kaon-loop contribution discussed in
Sec.~\ref{sec:nEDM}. For instance, if we take $\Lambda$ to be the
$\rho$-meson mass, the strangeness contribution is found to be
suppressed by 30~\%, while with $\Lambda=m_\Sigma$, it is enhanced by
about 40~\%. Hence, much detailed analyses of the chiral-loop
contributions as well as the estimate of the tree-level counterterms are
required in order to reduce the uncertainty.

Compared to the results obtained with the QCD sum rules
\cite{hep-ph/0010037, Hisano:2012sc}, our results here predict
relatively large neutron EDM. Indeed, a latest calculation
conducted in Ref.~\cite{Hisano:2012sc} gives\footnote{Equations
(\ref{dn_sum_rule}) and (\ref{dnPQ_sum_rule}) also contain the
contribution of quark EDMs; $0.79 d_d-0.20d_u$ with $d_u$ and $d_d$ the
EDMs of up and down quarks, respectively. Further, the
numerical values presented here are in fact different from those
in Ref.~\cite{Hisano:2012sc} by nearly a factor of two. The difference
stems from the use of different values for the quark condensate; 
$\langle \bar{q}q\rangle=-(225~{\rm MeV})^3$ in
Ref.~\cite{Hisano:2012sc}. }
\begin{equation}
 d_n=e(8.2\times 10^{-17}\bar{\theta}~[{\rm cm}]-0.30\tilde{d}_u
+0.30\tilde{d}_d-0.01\tilde{d}_s),
\label{dn_sum_rule}
\end{equation} 
and in the presence of the PQ symmetry, it reduces to 
\begin{equation}
 d_n\vert_{\rm PQ}=e(0.30\tilde{d}_u+0.59\tilde{d}_d).
\label{dnPQ_sum_rule}
\end{equation}
These values are found to be smaller than ours by ${\cal O}(1)$
factors\footnote{ 
Relatively larger values follow from the results in
Ref.~\cite{hep-ph/0010037}, while they are still smaller than those
obtained in the present work. Note that the expressions for the CEDM
contributions in Refs.~\cite{hep-ph/0010037} and \cite{Hisano:2012sc}
are different from each other. The difference is pointed out just below
Eq.(70) in Ref.~\cite{Hisano:2012sc}. 
}. 
The expression tells us that the strangeness contribution
vanishes when you exploit the method of QCD sum rules, where the sea
quark effects show up only at the higher orders. The same
consequence is obtained if you use the results in
Ref.~\cite{hep-ph/0010037}. 
With the present knowledge, we are not able to conclude which method is
appropriate for the calculation of the neutron EDM, since both of them may
have substantial uncertainty.

\section{Conclusion}
\label{sec:conclusion}

We have calculated the neutron electric dipole moment in the presence of
the CP-violating operators up to dimension five based on the chiral
perturbation theory. In the calculation, we extract the nucleon matrix
elements of scalar-type quark operators, $S_q$, from the recent lattice
results, while those of the dipole-type quark-gluon operators, $D_q$,
are evaluated by using the method of the QCD sum rules. Especially, we
focus on the strange CEDM contribution to the neutron EDM. Although the
lattice QCD simulations tell us that the strangeness content in nucleon 
is small, it is found the contribution may not be suppressed. 

There are two origins of the theoretical error in our calculation; one
is unknown counterterms in the effective Lagrangian which contribute to
the neutron EDM at tree level, and the other is the evaluation of $D_q$
by using the method of QCD sum rules. The former is somewhat inevitable
until one fixes all of the low-energy constants for the
counterterms. The latter is, on the other hand, expected to be improved
if the higher order calculation is conducted with unknown condensates
determined from other studies. Further, the lattice QCD simulations
might compute $D_q$ directly. It allows us to make a prediction of
neutron EDM much precisely with error estimated more systematically.

At present, we conclude that the contribution of strange quark CEDM may be
still significant, and therefore it might offer a sensitive probe for
the CP-violating interactions in physics beyond the Standard Model.

\section*{Acknowledgments}

We would like to thank M. Tanabashi for useful comments.
This work is supported by Grant-in-Aid for Scientific research from
the Ministry of Education, Science, Sports, and Culture (MEXT), Japan,
No. 20244037, No. 20540252, No. 22244021 and No.23104011 (JH), and
also by World Premier International Research Center Initiative (WPI
Initiative), MEXT, Japan. The work of NN is supported by
Research Fellowships of the Japan Society for the Promotion of Science
for Young Scientists.

\section*{Appendix}
\appendix

\section{Baryon matrix elements}
\label{baryon_mat}

In this section, we derive formulae for the baryon matrix elements given
in Eq.~(\ref{PCAC1}). To that end, we first rewrite ${\cal A}$ in terms
of the SU(3) generators: 
\begin{equation}
 {\cal A}=\frac{1}{3}({\cal A}_u+{\cal A}_d+{\cal A}_s)\1+
({\cal A}_u-{\cal A}_d)T^3
+\frac{1}{\sqrt{3}}({\cal A}_u+{\cal A}_d-2{\cal A}_s)T^8~.
\end{equation}
Then, by using the anticommuting relations
\begin{equation}
 \{T^A, T^B\}=\frac{1}{3}\delta_{AB}+d_{ABC}T^C~,
\end{equation}
we obtain
\begin{align}
 \langle B^B&\vert\overline{Q}\{T^A,{\cal A}
\}Q\vert B^C\rangle \nonumber \\
&=
\frac{2}{3}\langle B^B\vert \overline{Q}(
{\cal A}_u+{\cal A}_d+{\cal A}_s) T^A Q\vert B^C\rangle
+\frac{1}{3}\delta_{A3}\langle B^B\vert\overline{Q}
({\cal A}_u-{\cal A}_d) Q\vert B^C\rangle \nonumber \\
&+d_{A3D}\langle B^B\vert \overline{Q}(
{\cal A}_u-{\cal A}_d) T^D Q\vert B^C\rangle
+\frac{1}{3\sqrt{3}}\delta_{A8}\langle B^B\vert \overline{Q}(
{\cal A}_u+{\cal A}_d-2{\cal A}_s)  Q\vert B^C\rangle
\nonumber \\
&+\frac{1}{\sqrt{3}}d_{A8D}\langle B^B\vert \overline{Q}(
{\cal A}_u+{\cal A}_d-2{\cal A}_s) T^D Q\vert B^C\rangle~.
\label{138}
\end{align}
Here, we define $d_{ABC}$ as $d_{ABC}\equiv 2{\rm Tr}[\{T^A, T^B\}T^C]$.

Next, we express the baryon matrix elements in the above equation in
terms of the baryon matrix field $B$ defined in
Eq.~(\ref{baryon_mat_field}). Since there are two different
ways of combining three matrices to form an SU(3) covariant form, we
parametrize the baryon matrix elements as
\begin{align}
 \langle B^B\vert \overline{Q}(
{\cal A}_u+{\cal A}_d+{\cal A}_s) T^A Q\vert B^C\rangle
\overline{B^B}B^C
&\equiv X_1 {\rm Tr}(\overline{B}T^AB)+Y_1{\rm Tr}(\overline{B}BT^A)~,
\nonumber\\
 \langle B^B\vert \overline{Q}(
{\cal A}_u-{\cal A}_d) T^A Q\vert B^C\rangle
\overline{B^B}B^C
&\equiv X_3 {\rm Tr}(\overline{B}T^AB)+Y_3{\rm Tr}(\overline{B}BT^A)~,
\nonumber\\
 \langle B^B\vert \overline{Q}(
{\cal A}_u+{\cal A}_d-2{\cal A}_s) T^A Q\vert B^C\rangle
\overline{B^B}B^C
&\equiv X_8 {\rm Tr}(\overline{B}T^AB)+Y_8{\rm Tr}(\overline{B}BT^A)~.
\label{XY}
\end{align}
Further, considering the combination of two matrices to form an $SU(3)$
singlet, we obtain
\begin{align}
 \langle B^B\vert \overline{Q}(
{\cal A}_u-{\cal A}_d) Q\vert B^C\rangle
\overline{B^B}B^C
&\equiv Z_3 {\rm Tr}(\overline{B}B)~,
\nonumber\\
 \langle B^B\vert \overline{Q}(
{\cal A}_u+{\cal A}_d-2{\cal A}_s)  Q\vert B^C\rangle
\overline{B^B}B^C
&\equiv Z_8 {\rm Tr}(\overline{B}B)~.
\label{Z}
\end{align}
These parameters can be written in terms of proton matrix elements.
Then, using $S_q$ and $D_q$ defined in Eq.(\ref{nucl_mat_def}), we
express the parameters in Eqs.~(\ref{XY}) and (\ref{Z}) as
\begin{align}
 X_1&=-(m_u\theta_u +m_d\theta_d+m_s\theta_s)
(S_u-S_d)-\frac{1}{2}(\tilde{d}_u+\tilde{d}_d+\tilde{d}_s)
(D_u-D_d)~,\nonumber \\
 X_3&=-(m_u\theta_u -m_d\theta_d)
(S_u-S_d)-\frac{1}{2}(\tilde{d}_u-\tilde{d}_d)
(D_u-D_d)~,\nonumber \\
 X_8&=-(m_u\theta_u +m_d\theta_d-2m_s\theta_s)
(S_u-S_d)-\frac{1}{2}(\tilde{d}_u+\tilde{d}_d-2\tilde{d}_s)
(D_u-D_d)~,\nonumber \\
 Y_1&=-(m_u\theta_u +m_d\theta_d+m_s\theta_s)
(S_s-S_d)-\frac{1}{2}(\tilde{d}_u+\tilde{d}_d+\tilde{d}_s)
(D_s-D_d)~,\nonumber \\
 Y_3&=-(m_u\theta_u -m_d\theta_d)
(S_s-S_d)-\frac{1}{2}(\tilde{d}_u-\tilde{d}_d)
(D_s-D_d)~,\nonumber \\
 Y_8&=-(m_u\theta_u +m_d\theta_d-2m_s\theta_s)
(S_s-S_d)-\frac{1}{2}(\tilde{d}_u+\tilde{d}_d-2\tilde{d}_s)
(D_s-D_d)~,\nonumber \\
 Z_3&=-(m_u\theta_u -m_d\theta_d)
(S_u+S_d+S_s)-\frac{1}{2}(\tilde{d}_u-\tilde{d}_d)
(D_u+D_d+D_s)~,\nonumber \\
 Z_8&=-(m_u\theta_u +m_d\theta_d-2m_s\theta_s)
(S_u+S_d+S_s)-\frac{1}{2}(\tilde{d}_u+\tilde{d}_d-2\tilde{d}_s)
(D_u+D_d+D_s)~.
\label{XYZ}
\end{align}

\section{Quark propagators and correlators of the background fields}
\label{propagator_correlator}

In the OPE calculation carried out in Sec.~\ref{Dq}, we need to obtain
the quark propagators as well as the correlators of the quark/gluon
background fields in the presence of the interaction in
Eq.~(\ref{external}). The quark propagators are defined as follows: 
\begin{equation}
 [S^q_{ab}(x) ]_{\alpha\beta}
\equiv
\langle\Omega|T
\left[q_{a\alpha}(x)\bar{q}_{b\beta}(0) \right]| 
\Omega\rangle_{J}~, 
\end{equation}
where $\alpha$ and $\beta$ denote spinor indices. 
Furthermore, we perturbatively expand the propagators as
\begin{equation}
  \left[S^q_{ab}(x) \right]_{\alpha\beta}
=
 \left[S^{q(0)}_{ab}(x) \right]_{\alpha\beta}+
\chi^q_{a\alpha}(x)\bar{\chi}^q_{b\beta}(0)
+
\left[S^q_{ab}(x) \right]_{\alpha\beta}|_{g}+ \dots~.
\label{propagator}
\end{equation}
The first term is the free
propagator, and the second term describes the correlator of the quark
background fields, with $\chi^q_{a\alpha}(x)$ a classical Grassmann field
which indicates the quark background field.
The third term represents the propagator which includes one gluon.
Let us evaluate these terms in $x$-space.
The first term, $[S^{q(0)}_{ab}(x)
]_{\alpha\beta}$, is readily evaluated as
\begin{align}
 S^{q(0)}_{ab}(x)&=\frac{i\delta_{ab}}{2\pi^2}\frac{\Slash{x}}{(x^2)^2},
\label{propagator0}
\end{align}
where we neglect the quark masses since their contribution only
appears in the higher order operators. 

Next, we evaluate the third term in Eq.~(\ref{propagator}). In this
calculation, it is convenient to exploit the Fock-Schwinger gauge
\cite{Novikov:1983gd} for the gluon field. In this gauge, the gluon
field is subjected to the following gauge fixing condition:
\begin{equation}
 x^\mu G_\mu(x)= 0~,
\end{equation}
where $G_\mu (x)\equiv G^A_\mu (x)T^A$ is the gluon field.
Then, it is expanded by its field strength tensor such that
\begin{equation}
 G_\mu(x)=
\frac{1}{2\cdot 0 !} x^{\nu} G_{\nu\mu}(0)
+
\frac{1}{3\cdot 1 !} 
x^\alpha x^{\nu} (D_\alpha  G_{\nu\mu}(0))
+
\frac{1}{4\cdot 2 !} 
x^\alpha x^\beta x^{\nu} (D_\alpha D_\beta G_{\nu\mu}(0))
+
\cdots.
\label{Aexpansion}
\end{equation}
By using the expression, the gauge covariant form
of the propagators is obtained as follows:
\begin{align}
 S^q_{ab}(x)|_g=
&-\frac{g_s}{32\pi^2}\biggl[
\frac{i}{x^2}\{\Slash{x},G_{ab}\cdot \sigma\}
+4J_q
\frac{\Slash{x}G_{ab}\cdot \sigma\Slash{x}}{(x^2)^2}~
\biggr],
\label{gluon_propagator}
\end{align}
with $G_{ab}^{\mu\nu}=G^{A\mu\nu}T^A_{ab}$. 
Here we keep only the first order terms with respect to the
external fields, $J_q$. 

Also, we translate the quark and gluon background
fields into their condensates. The correlation function of quark
background fields, $\chi^q_{a\alpha}(x)\bar{\chi}^q_{b\beta}(0)$, is
related with the quark condensate as
\begin{equation}
 \chi^q_{a\alpha}(x)\bar{\chi}^q_{b\beta}(0)=
\langle \Omega|T[ q_{a\alpha}(x)\bar{q}_{b\beta}(0) ]|\Omega
\rangle.
\end{equation}
By using the Fierz identities and carrying out the short-distance
expansion of the quark field,
\begin{equation}
 q(x)=q(0)+x^\mu D_\mu q(0)+\dots~,
\end{equation}
we obtain
\begin{align}
  \chi^q_{a\alpha}(x)\bar{\chi}^q_{b\beta}(0)
=&-\frac{\delta_{ab}}{12}
\langle\bar{q}q\rangle
[\1
-\frac{i}{4}\Slash{x}J_q m^2_0
]_{\alpha\beta}~.
\label{chichi}
\end{align}
In general, the vacuum condensate in the presence of external fields,
$\langle \bar{q}q \rangle\vert_J$, is different from ordinary one,
$\langle \bar{q}q \rangle$\cite{Jin:1993nn}. However, the
susceptibilities of the condensates to the external fields only appear
in the higher order of the OPE, therefore, we neglect them since in our
calculation we just deal with the leading order contributions. 

In addition, we need the interaction part of the quark and
gluon background fields,  
\begin{equation}
 g_s\chi^q_{a\alpha}(x) \bar{\chi}^q_{b\beta}(0)
G_{\mu\nu}^A=
\langle \Omega |T[g_s q_{a\alpha}(x)G_{\mu\nu}^A \bar{q}_{b\beta}(0)]
|\Omega\rangle,
\end{equation}
and it leads to the following equation:
\begin{align}
 & g_s\chi^q_{a\alpha}(x) \bar{\chi}^q_{b\beta}(0)
G_{\mu\nu}^A=
 \frac{1}{192}T^A_{ab}
m_0^2\langle\bar{q}q\rangle
(
\sigma_{\mu\nu})_{\alpha\beta}.
\label{chichig}
\end{align}

{}

\end{document}